# A Gaussian Thermionic Emission Model for Analysis of Au/MoS$_2$ Schottky Barrier Devices

*Calvin Pei Yu Wong,*[1,2,3,*] *Cedric Troadec,*[2] *Andrew T. S. Wee*[1,2]*, Kuan Eng Johnson Goh,*[2,3,*]

[1]*NUS Graduate School for Integrative Sciences and Engineering, National University of Singapore, Centre for Life Sciences, #05-01, 28 Medical Drive, Singapore 117456, Singapore*

[2]*Institute of Materials Research and Engineering, A\*STAR (Agency for Science, Technology and Research), 2 Fusionopolis Way, #08-03 Innovis, Singapore 138634, Singapore*

[3]*Department of Physics, Faculty of Science, National University of Singapore, 2 Science Drive 3, Singapore 117551, Singapore*

\*Authors to whom correspondence should be addressed: calvin_wong@imre.a-star.edu.sg; kejgoh@yahoo.com

**Abstract:**

Schottky barrier inhomogeneities are expected at the metal/TMDC interface and this can impact device performance. However, it is difficult to account for the distribution of interface inhomogeneity as most techniques average over the spot-area of the analytical tool (e.g. few hundred micron squared for photoelectron-based techniques), or the entire device measured for electrical current–voltage (*I-V*) measurements. Commonly used models to extract Schottky barrier heights neglect or fail to account for such inhomogeneities, which can lead to the extraction of incorrect Schottky barrier heights and Richardson constants that are orders of magnitude away from theoretically expected values. Here, we show that a gaussian modified thermionic emission model gives the best fit to experimental temperature dependent current–voltage (*I-V-T*) data of van der Waals Au/*p*-MoS$_2$ interfaces and allow the deconvolution of the Schottky barrier heights of the defective regions from the pristine region. By the inclusion of a gaussian distributed Schottky barrier height in the macroscopic *I-V-T* analysis, we demonstrate that interface inhomogeneities due to defects are deconvoluted and well correlated to the impact on the device behavior across a wide temperature range from room temperature of 300 K down to 120 K. We verified the gaussian thermionic model across two different types of *p*-MoS$_2$ (geological and synthetic), and finally compared the macroscopic Schottky barrier heights with the results of a nanoscopic technique, ballistic hole emission microscopy (BHEM). The results obtained using BHEM were consistent with the pristine Au/p-MoS$_2$ Schottky barrier height extracted from the gaussian modified thermionic emission model over hundreds of nanometers. Our findings show that the inclusion of Schottky barrier inhomogeneities in the analysis of *I-V-T* data is important to elucidate the impact of defects (e.g. grain boundaries, metallic impurities, etc.) and hence their influence on device behavior. We also find that the Richardson constant, a material specific constant typically treated as merely a fitting constant, is an important parameter to check for the validity of the transport model.

# I. Introduction

Schottky barrier inhomogeneities are present at metal/transition metal dichalcogenide (TMDC) interface due to potential fluctuations and defects in the material and this can impact the device performance [1–4]. However, the most commonly used thermionic emission transport model modified with a simple ideality factor correction [5] does not account sufficiently for non-ideal effects such as inhomogeneities, often leading to the extraction of an apparent Schottky barrier height (SBH) convoluted with other factors such as defects (inhomogeneity) and temperature. The extracted apparent SBH from the simple model does not represent the true band alignment at the metal/semiconductor interface and is counterproductive to the correct understanding of the energetics of the interface. Several other models have been proposed to account for these non-ideal effects in the current–voltage (*I-V*) behavior of Schottky barrier devices [1–8]. In this paper, we compare the effectiveness of four types of commonly used transport models to extract correct values of the SBH and the material specific Richardson constant of $MoS_2$. We verified the SBH extracted from the transport models against the SBH obtained from ballistic hole emission microscopy (BHEM) [9–11], which a direct measurement method for the SBH at the nanoscale, and verified the Richardson constants extracted from the models with theoretical calculated values based on the electron (hole) effective mass of $MoS_2$ [12]. Finally, we compared the SBH and Richardson constants across two types of $MoS_2$ (geological and synthetic) from different sources and show that the SBH and Richardson constants are similar across the two $MoS_2$ devices if analyzed using the correct model.

The transport of thermally activated carriers across a typical metal/semiconductor Schottky interface where the semiconductor is lightly doped ($\sim 10^{15}$ to $10^{17}$ cm$^{-3}$) is given by the ideal thermionic emission model (Eq. 1) [5].

$$I = I_S \left[\exp\left(\frac{qV}{kT}\right) - 1\right] \quad (1)$$

where

$$I_S = \left[AA^{**}T^2 \exp\left(-\frac{q\phi}{kT}\right)\right] \quad (2)$$

and $q$ is the electric charge, $V$ the voltage applied across the diode, $k$ the Boltzmann constant, $T$ the absolute temperature, $A$ the area of the diode, $A^{**}$ the effective Richardson constant of the semiconductor and $\phi$ the Schottky barrier height (SBH). An important requirement of this model is that the tunneling of carriers across the Schottky barrier is negligible, which is valid when the semiconductor is lightly doped (~$10^{15}$ to $10^{17}$ cm$^{-3}$) such that the band banding is gradual and the Schottky barrier width is wide. The accurate extraction of the SBH depends greatly on the successful determination of $I_S$ from the forward bias slope of the experimental $I$-$V$ curves (Eq. 1). However, the thermionic emission model does not fit well to experimental $I$-$V$ curves at low temperatures, and the modified (non-ideal) thermionic emission models [6,7], the thermionic field emission (thermally assisted tunneling) model or the generation-recombination model are often used instead [13,14]. While these modified models seem to fit certain experimental datasets well, the numbers that have been extracted have not always been reliable. For instance, negative SBHs have been reported [15,16], but negative SBHs have no physical meaning and show that the models are not suitable for these specific devices, and other material specific constants such as the Richardson constants are orders of magnitude away from the theoretically derived values [6,7], Hence, it is not trivial to identify a correct model to accurately extract the SBH of metal/semiconductor Schottky interfaces, especially if the measurement is done at a specific, or a small range of temperatures, as the effect of temperature can be convoluted into the extracted SBH.

In this paper, we investigate the use of four well established methods for extracting Schottky barrier heights as a function of temperature from a weakly-interacting van der Waals Au contact to a layered $MoS_2$ crystal. The epitaxial growth of Au on $MoS_2$ allows a clean and abrupt interface for this material system [17], similar to recent reports of fabricating a van der Waals metal/TMDC interface [18,19], making it an ideal model system for this study. Four methods are used to extract SBH from temperature dependent *I-V* (*I-V-T*) curves: **Method 1:** the standard Richardson plot $\ln(I_s/T^2)$ vs $1/T$ [5], **Method 2:** the modified Richardson (Hackam and Haarop) plot $\ln(I_s/T^2)$ vs $1/nT$ [8] and **Method 3:** the modified Richardson (Bhuiyan) plot $n\ln(I_s/T^2)$ vs $1/T$ [6], From this comparison, we show that none of these three methods provided a satisfactory fit, while a fourth method, **Method 4:** the gaussian thermionic emission model [4], provides the best fit across a wide temperature regime of 120 K to 300 K across two different types (geological and synthetic) of $MoS_2$ samples. To date, although the gaussian thermionic model has shown much success in fitting experimental data to metal/$MoS_2$ Schottky devices [20,21], this model has not been compared directly across samples of different origins and explicitly verified with a complementary technique, such as ballistic electron (hole) emission microscopy [9,10,22], which is a direct measurement of the nanoscale unbiased SBH. The aim of this paper is to review and compare these analysis methods in the context of a clean van der Waals epitaxial contact to a layered material, and show that using the inadequate model in the analysis of temperature dependent *I-V* data can yield Schottky barrier height values that are misleading by up to an order of magnitude and counterproductive to the understanding of the Schottky interface. Importantly, we show that the effective Richardson constant ($A^{**}$), typically treated as a fitting constant is an important parameter to cross check the validity of the transport model. As field effect transistors based on 2 dimensional TMDCs such as $MoS_2$ and $WS_2$ have been shown to behave as Schottky barrier transistors [23–25], where the Schottky barrier at the metal/semiconductor contact is modulated by the gate bias, the correct analysis of the Schottky barrier is crucial to the understanding of the subthreshold behavior of these 2D transistors, especially in the presence of defects [26–28].

**II.     Methods to extract the Schottky Barrier Height (SBH)**

The thermionic emission model, Eq. 1, predicts that for $V > 3kT/q$, a plot of ln *I* against *V* will be linear with a slope of 1 and its intercept at $V = 0$ will give $I_s$. From $I_s$, the Schottky barrier height can be extracted. A direct reading of $I_s$ from the experimental *I-V* curve is typically not used as the experimental reverse biased saturation current, as it also contains the image force lowering and other minority carrier effects [5]. However, the thermionic emission model is inadequate for a realistic metal/semiconductor interface especially at low temperatures. To account for the non-ideal transport mechanisms, and series resistance in real devices, the ideality factor $n$ and the series resistance $R_s$, are empirically added to the model [5,29] and Eq. 1 becomes:

$$I = I_S \left[ \exp\left(\frac{q(V - IR_s)}{nkT}\right) - 1 \right] \quad (3)$$

Using the modified thermionic emission model, diode parameters such as the ideality factor *n* and barrier height $\phi$ can be plotted and were found to be dependent on the temperature. At low temperatures, the thermionic emission model does not fit well to the experimental data and $n$ increases greatly beyond 1, signifying non-ideal transport. While the non-ideal transport has been attributed to additional current contributions from thermally assisted tunneling (thermionic field emission) across the Schottky barrier, generation-recombination current in the depletion region and image force lowering, it is not clear how the empirically modified thermionic emission model can account for these effects. From these modifications, a few versions of the Richardson plot will be analyzed.

A. **Method 1: Ideal Richardson plot (ln $I_s/T^2$ vs $1/T$)**

This is the simplest method and is derived directly from the saturation current term of the thermionic emission model, Eq. 2. When temperature dependent plots can be obtained, a plot of $\ln(I_S/T^2)$ vs. $1/T$, called a Richardson plot, will be a straight line where the slope and intercept at $1/T = 0$ will give $\phi$, and $A^{**}$ respectively. The empirically added ideality factor is not included in Eq. 2, hence this method is the ideal Richardson plot analysis.

B. **Method 2: Ideality factor modified Richardson plot I (ln $I_s/T^2$ vs $1/nT$)**

To account for effects that cause deviations from ideal ($n = 1$) behavior, such as the image force and surface charges, which they argued to be also present at zero bias, Hackam and Harrop proposed a modified Richardson plot from Eq. 3 to include the ideality factor in the $I_s$ term [8]. The forward current now looks:

$$I = \left[AA^{**}T^2 \exp\left(-\frac{q\phi}{nkT}\right)\right]\left(\frac{q(V-IR_s)}{nkT} - 1\right) \quad (4)$$

The addition of $n$ to the $I_s$ term in Eq. 4 now gives a modified Richardson plot Eq. 5 from which the SBH can be extracted from the gradient of the straight line and $A^{**}$ from the $y$-intercept.

$$\ln\frac{I_S}{T^2} = \ln AA^{**} - \frac{q\phi}{nkT} \quad (5)$$

C. **Method 3: ideality factor modified Richardson plot II ($n\ln I_s/T^2$ vs $1/T$)**

Bhuiyan, Martinez and Esteve found that the Hackam and Harrop model does not work for them due to the presence of a strongly temperature dependent SBH and ideality factor measured [6] and that the $A^{**}$ extracted from using the Hackam and Harrop method is too large. Hence, they empirically proposed Eq. 6:

$$I = \left[AA^{**}T^2 \exp\left(-\frac{q\phi}{nkT}\right)\right]\left[\exp\left(\frac{q(V-IR_s)}{kT}\right)-1\right] \quad (6)$$

Following their modification, the modified Richardson plot now reads:

$$n \ln \frac{I_S}{T^2} = \ln AA^{**} - \frac{q\phi}{kT} \quad (7)$$

### D. Method 4: Inhomogeneous Gaussian barrier modified Richardson plot

Two different inhomogeneous Schottky barrier models have been proposed independently by Werner and Güttler [3,4], and Tung and coworkers [2,30]. Werner and Güttler used a gaussian approximation of the SBH distribution to account for the potential fluctuations at the interface, while Tung used a generalized model. While Tung's model is more rigorous, Werner and Gutter's model is simpler and can be placed into the context of BHEM measured SBHs. Hence in this paper, we focused on the Werner and Güttler model of gaussian SBH [4], which is given by:

$$\phi^{app} = \Phi - \frac{\sigma^2}{2kT/q} \quad (8)$$

where $\phi^{app}$ is the apparent Schottky barrier height obtained as a result of the convolution of the gaussian distributed SBH with temperature in the thermionic emission model, $\Phi$ is the mean Schottky barrier height and $\sigma$ is the standard deviation of the gaussian distribution. To obtain the $\sigma$ of the Gaussian, a plot of $\phi^{app}$ against $1/T$ can be used. The gaussian standard deviations ($\sigma$) extracted from Eq. 8 can then be used to correct for the gaussian distributed SBH to obtain the gaussian corrected Richardson plots Eq. 9. Here, a plot of $\ln\left(\frac{I_S}{T^2}\right) - \frac{q^2\sigma^2}{2k^2T^2}$ against $\frac{1}{T}$ will give the $A^{**}$ in the intercept and $\Phi$ in the gradient.

$$\ln\left(\frac{I_S}{T^2}\right) = \ln AA^{**} - \frac{q\Phi}{kT} + \frac{q^2\sigma^2}{2k^2T^2} \quad (9)$$

The temperature dependence of the ideality factor and SBH, initially viewed as empirical inconsistencies in many experiments, is now well explained by Werner and Güttler to arise from the inhomogeneous SBH and that capacitance-voltage (*C-V*) measurements give Φ. In our experiments, we did not perform *C-V* measurements as capacitance measurement is not typically used in the operation of devices, but the current as a function of applied voltage (*I-V*) is used and is more common for analysis. Werner and Güttler also showed that for lightly doped ($10^{15}$ to $10^{17}$ cm$^{-3}$) semiconductors, thermionic emission dominates carrier transport, even at low temperatures down to 77 K.

We demonstrate in our Au/MoS$_2$ sample that by using a gaussian distributed SBH to account for these inhomogeneities, a more reliable value of the SBH can be obtained. We verify the gaussian thermionic emission model systematically using two different types of MoS$_2$ (geological and synthetic crystals, from different suppliers) [31,32] by performing temperature dependent current voltage measurements (*I-V-T*). We show that the *A*\*\*, an important material specific constant dependent only on the electron effective mass, though commonly treated as merely a fitting constant, can be a useful parameter to cross-check the validity of the model. Therefore, it is important to obtain Richardson plots for these devices. To further validate the use of the gaussian thermionic emission model, we compared the extracted mean SBH (Φ) with a complementary technique, ballistic hole emission microscopy (BHEM) and show that the SBH values obtained are identical within error limits across the two different samples and complementary techniques. We propose that the gaussian thermionic emission model gives a more accurate representation of the real Schottky interface and our results can be used to reconcile the conflicting reports on SBH in the literature and allow *I-V-T* analysis to yield more in depth understanding of the interface.

### III. Experimental Design

Figure 1 shows a schematic of the fabricated Au/MoS$_2$ Schottky diode and the corresponding scanning electron microscope (SEM) image. We used the bulk MoS$_2$ crystal as the device material to allow clean shadow mask fabrication and the metal/semiconductor interface formed with the top layer can provide the basic understanding of metal contacts to layered semiconductor devices. Ti was chosen as the ohmic contact as it is a commonly used material for ohmic contacts to MoS$_2$. We chose Au as the Schottky contact as it is a high work function metal that is not expected to form ohmic contacts with MoS$_2$ without interface modification, and it is also a commonly used contact material in the literature due to its inertness in the ambient environment.

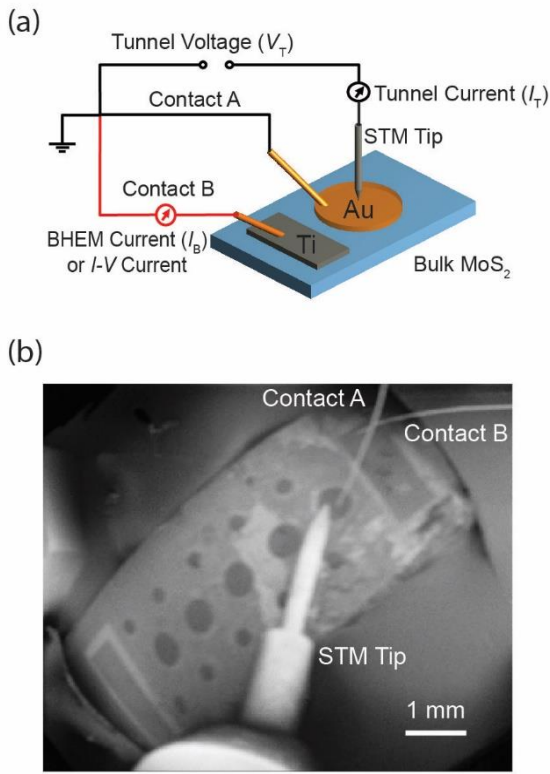

Figure 1 (a) Schematic of the Au/MoS$_2$ device. Contacts A and B are used to perform *I-V* measurements and the scanning tunneling microscope (STM) tip is added for BHEM measurements (b) the corresponding scanning electron micrograph showing actual device during BHEM measurements. Imaging conditions: 0.1 kV acceleration voltage, 20 pA current.

## IV. Experimental

The *p*-type geological $MoS_2$ crystal was obtained from Ward's Science [31], the *p*-type synthetic $MoS_2$ crystal (intrinsic) was obtained from 2D semiconductors [32]. The ohmic contacts to the $MoS_2$ crystals were deposited in a high vacuum e-beam evaporator system (AJA International) after cleaving off the top surface using sticky tape to obtain a fresh surface for the evaporation of Ti(5 nm)/Au(80 nm) at a base pressure of $1 \times 10^{-8}$ mbar to form ohmic contacts. After deposition of the ohmic contacts, the $MoS_2$ was transferred ex-situ to a thermal evaporator system equipped with an annealing stage (R-DEC) for the Schottky contact deposition. First, the $MoS_2$ crystal was annealed at 350 °C for 2.5 h to improve the quality of the ohmic contacts by removing physisorbed material at the interface or to promote an interface reaction of Ti with the $MoS_2$ surface, and to outgas physisorbed contaminants on the surface of the $MoS_2$ crystal. Next, Au (15 nm) was thermally evaporated onto the clean surface of the $MoS_2$ crystal through a shadow mask at the rate of 0.2 Å/s at approximately 50 °C at a base pressure of $1 \times 10^{-8}$ mbar. The larger circular Au devices are 500 μm in diameter and the smaller circular Au devices are 250 μm in diameter, while the rectangular Ti/Au ohmic contacts on each sides of the substrate are $0.7 \times 7$ mm². Finally, the samples were transferred ex-situ to an UHV Nanoprobe system (Omicron) which is an ultra-high vacuum (base pressure $1 \times 10^{-10}$ mbar) four probe STM system equipped with three standard STM probes for contacted *I-V* measurements, one atomic resolution capable STM probe and a Zeiss Gemini SEM imaging column for accurate positioning of the probes. The devices were measured without further annealing to prevent modification of the as deposited interface. The manifestation of epitaxial Au(111) steps in the overgrowth Au cap layer (Figure 6) provides an indication of good interface cleanliness/quality of this preparation method.

## V. Results

***I-V-T* measurements.** Figure 2 shows the *I-V-T* measurements of Au/geological MoS$_2$ and the Au/synthetic MoS$_2$ Schottky diodes. From the *I-V-T* characteristics of the devices, we observe the typical rectifying behavior of a *p*-type Schottky diode with low leakage at positive bias and current turn on at negative biases and hence can conclude that our MoS$_2$ devices are *p*-type. A shunt conduction pathway which could arise from conduction through defective regions with lower resistance is present for the geological MoS$_2$ crystal. The synthetic MoS$_2$ device shows a typical diode *I-V* characteristic with low leakage under reverse bias, turn on at threshold followed by monatomic rise in *I*, with no defect dominated shunt at low biases, from which we can conclude that the synthetic MoS$_2$ crystal is cleaner. The presence of a high series resistance complicates the analysis due to voltage drop across the series resistance, but can be corrected using the Werner method [33] (Supplemental Information, Figure S1). We fitted all the individual *I-V* curves to the standard ideality factor modified thermionic emission model (Eq. 3), to extract the Schottky barrier heights ($\phi$) and ideality factors ($n$) as a function of temperature and plotted the extracted values for the geological device in Figure 2b and the synthetic device in Figure 2d. $A^{**} = 400\ 000$ A m$^{-2}$ K$^{-2}$ was used as the theoretical effective Richardson constant of *p*-MoS$_2$,[1] Figure 2b shows that the temperature dependent SBHs extracted from both MoS$_2$ substrates are inversely proportional to temperature and the ideality factors are proportional to temperature, similar to many reports in the literature. Furthermore, Figure 2d shows two regimes where the ideality factor scales differently as a function of temperature. Tung previously explained that this temperature dependence of the ideality factor is a signature of inhomogeneous SBH, similar to Method 4 of our analysis [1].

---

[1] This is calculated using the effective mass of holes at the valance band m* = 0.62 m$_0$ [51] and with a quantum correction factor of 0.5 following Cowley and Sze (1965). [12]

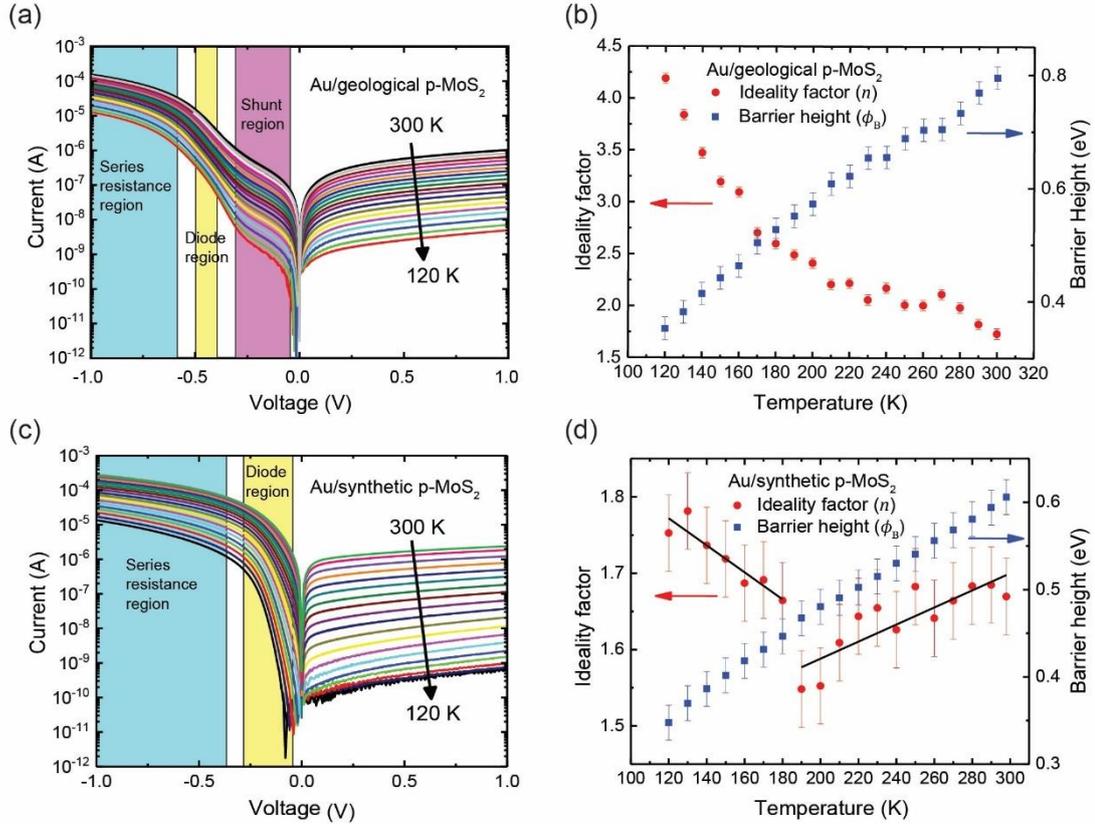

Figure 2 (a) Temperature dependent current–voltage (*I-V-T*) measurements of Au/geological *p*-MoS$_2$ from 300 K (room temperature) to 120 K. The device area is 0.196 mm$^2$. (b) Plots of extracted diode parameters for the Au/geological MoS$_2$ Schottky diode against temperature. Ideality factor (red, left axis) and barrier height (blue, right axis). (c) Temperature dependent current–voltage (*I-V-T*) measurements of Au/synthetic *p*-MoS$_2$ and (d) the extracted diode parameters for the Au/synthetic MoS$_2$ devices. Black lines are guides to the eye for showing two ideality factor ranges. The error bars of $n \pm 0.05$ and $\phi \pm 0.02$ eV are estimated from the error of range of fit of Eq. 3 to Figures 2a and c.

**Methods 1, 2 and 3.** To analyze the temperature dependence of the SBH and ideality factor, Figure 3 shows the Richardson plots analyzed using used Methods 1, 2 and 3. By plotting the temperature dependent Richardson plots, one can extract the effective Richardson constant and the SBH of the device from the gradient and the *y*-intercept respectively. However, the ideal Richardson plots and the ideality factor modified

Richardson plots [6,7] yield $A^{**}$ values which are orders of magnitude away from the theoretical values and expected values valid for the model, or give unrealistic SBH values. Table 1 shows the summary of the extracted $\phi$ and $A^{**}$ values using methods 1 2 and 3. Method 1 gives the poorest fit to the experimental data, especially for low temperatures below approximately 180 K, where there is a large deviation from the linear fit. If we constrain the fits above 180 K, for the geological MoS$_2$ sample, we extract $\phi_1 = 0.130$ eV; $A_1^{**} = 6.12 \times 10^{-6}$ Am$^{-2}$K$^{-1}$ and for the synthetic MoS$_2$ sample $\phi_1 = 0.216$ eV; $A_1^{**} = 0.198$ Am$^{-2}$K$^{-1}$. While the $\phi$ values are of a reasonable number, the $A^{**}$ values are orders of magnitudes away from the theoretical value calculated from the effective mass of holes at the valance band maxima of p-MoS$_2$ of $A^{**} = 400\,000$ Am$^{-2}$ K$^{-2}$. For Method 2, we extracted for the geological MoS$_2$ sample $\phi_2 = 0.884$ eV; $A_2^{**} = 5.43$ Am$^{-2}$K$^{-1}$ and for the synthetic MoS$_2$ sample $\phi_2 = 0.334$ eV; $A_2^{**} = 0.105$ Am$^{-2}$K$^{-1}$. For Method 3, for the geological MoS$_2$ sample $\phi_3 = 1.47$ eV; $A_3^{**} = 3.15 \times 10^6$ Am$^{-2}$K$^{-1}$ and for the synthetic MoS$_2$ sample $\phi_3 = 0.369$ eV; $A_3^{**} = 6.01 \times 10^{-6}$ Am$^{-2}$K$^{-1}$. For Methods 2 and 3, although a linear fit can be obtained, and reasonable numbers can sometimes be extracted (Method 3 yield an unreasonable SBH larger than the band gap of MoS$_2$ for the geological MoS$_2$ sample), the $A^{**}$ values still do not match with the theoretical values and are varying over orders of magnitudes. It is important to note that for the synthetic MoS$_2$ sample, although Methods 1, 2 and 3 gives SBH of about 0.3 eV, the $A^{**}$ values are incorrect over orders of magnitudes. Hence, we can conclude that it is important to consider both the SBH and the $A^{**}$ values together and that these models are inadequate, and a better model is required to explain the data.

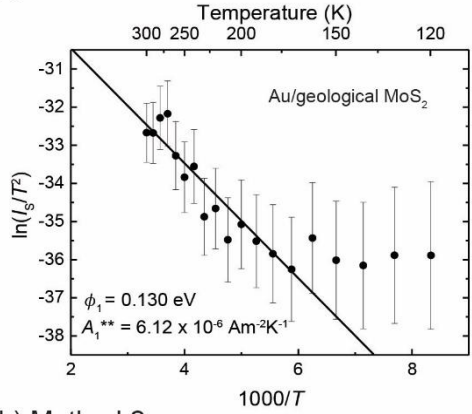
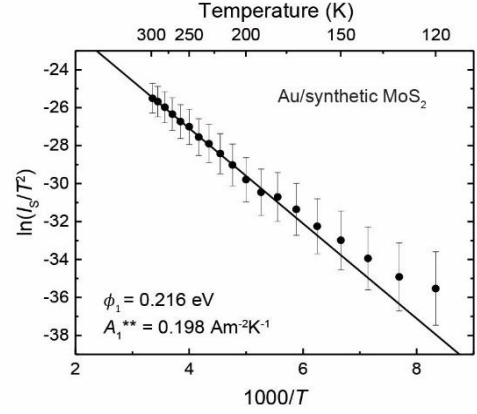
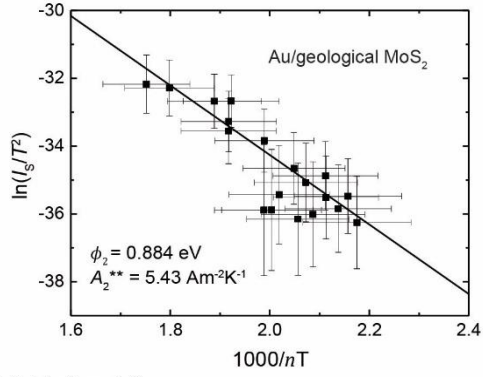
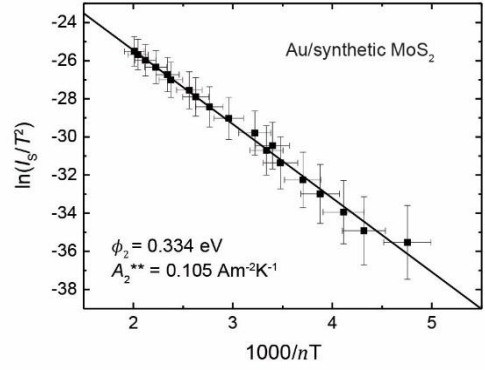
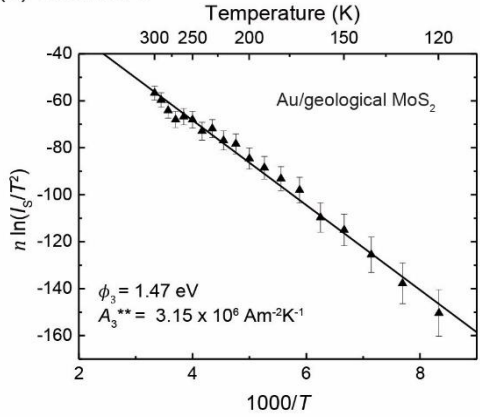
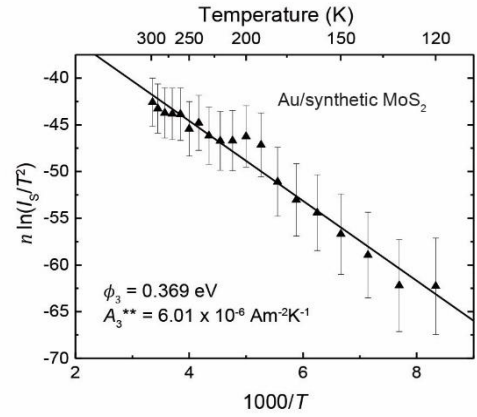

Figure 3 Richardson plots and their ideality factor modified Richardson plots variants for the Au/geological MoS$_2$ (left column) and the Au/synthetic MoS$_2$ (right column). (a) **Method 1:** Standard Richardson plot ln ($I_s/T^2$) against $1/T$. (b) **Method 2**: Modified ln ($I_s/T^2$) against $1/nT$ and (c) **Method 3**: modified $n$ ln ($I_s/T^2$) against $1/T$. Error bars are estimated with $\phi \pm 0.02$ eV and $n \pm 0.05$ from fitting errors to extract SBH and $n$.

Table 1. Extracted Schottky barrier heights and effective Richardson constants using Methods 1, 2 and 3.

|  | Au/geological MoS$_2$ | | Au/synthetic MoS$_2$ | |
| --- | --- | --- | --- | --- |
|  | $\phi$ (eV) | $A^{**}$ (A m$^{-2}$ K$^{-1}$) | $\phi$ (eV) | $A^{**}$ (A m$^{-2}$ K$^{-1}$) |
| **Method 1** | 0.130 | $6.12 \times 10^{-6}$ | 0.216 | 0.198 |
| **Method 2** | 0.884 | 5.43 | 0.334 | 0.105 |
| **Method 3** | 1.47 | $3.15 \times 10^{6}$ | 0.369 | $6.01 \times 10^{-6}$ |

**Method 4.** To further analyse the temperature dependence of the SBH, we use the inhomogeneous gaussian distributed Schottky barrier height model proposed by Werner and Güttler [4], which is given by Eq. 8. Plotting $\phi^{app}$ against $q/2kT$ in Figure 4 shows the presence of two different regimes where the two diode parameters dominate.

**Geological MoS$_2$.** Fitting Eq. 8 to Figure 4a shows that at temperatures above 195 ± 5 K, $\Phi_4^{B1} = 0.88 \pm 0.10$ eV dominates the device performance while at temperatures below 195 ± 5 K, $\Phi_4^{B2} = 1.18 \pm 0.14$ eV dominates the device performance. The gaussian standard deviations (σ) extracted from Figure 4a are then used to correct for the gaussian distributed SBH (Eq. 9) to obtain the gaussian corrected Richardson plots (Figure 4c). Fitting Eq. 9 to the experimental data in the respective temperature regime in Figure 4c, the mean SBH values obtained from the corrected Richardson plots of $\Phi_4^{B1} = 0.88 \pm 0.10$ eV and $\Phi_4^{B2} = 1.18 \pm 0.14$ eV are in excellent match with the values obtained in Figure 4a, $\Phi_4^{B1} = 0.88 \pm 0.10$ eV and $\Phi_4^{B2} = 1.17 \pm 0.14$ eV. The average Richardson constants extracted from the gaussian corrected Richardson plot of $A_4^{**} = 406\ 000 \pm 145\ 000$ Am$^{-2}$K$^{-1}$ is in good agreement with theoretical quantum corrected value of $A^{**} = 400\ 000$ Am$^{-2}$K$^{-1}$ [12], validating the feasibility of this method.

**Synthetic MoS$_2$.** Similarly for the synthetic MoS$_2$ crystal, Figure 4b shows that at high temperatures above 195 ± 5 K, the $\Phi_4^{B1} = 0.86 \pm 0.12$ eV dominates the device performance while at low temperatures, the $\Phi_4^{B2} = 0.65 \pm 0.08$ eV dominates the device performance. Plotting the gaussian modified Richardson plot in Figure 4d reveals a similar behaviour as the geological MoS$_2$ sample. The mean SBH values extracted from the Richardson plots of $\Phi_4^{B1} = 0.84 \pm 0.12$ eV and $\Phi_4^{B2} = 0.65 \pm 0.08$ eV are an excellent match with the values obtained in Figure 4b with a similar temperature crossover range of $195 \pm 5$ K. More importantly, the average $A_4^{**}$ extracted from the synthetic MoS$_2$ of $424\,000 \pm 23\,000$ Am$^{-2}$K$^{-1}$ is in excellent agreement with the $A_4^{**}$ of geological MoS$_2$ of $406\,000 \pm 145\,000$ Am$^{-2}$K$^{-1}$, and again with the theoretically derived values of $400\,000$ A m$^{-2}$ K$^{-2}$. These results provide crucial support for the validity of using the Gaussian modified SBH model in our samples.

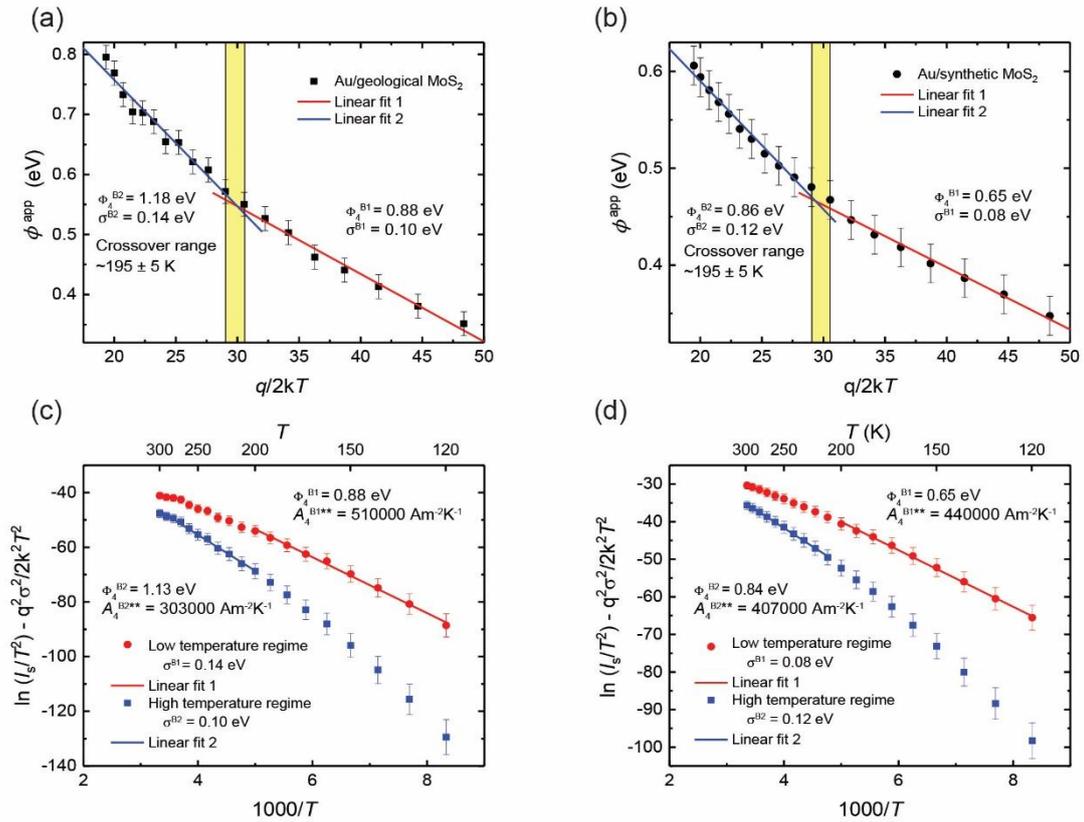

Figure 4 (a) Double gaussian plot of $\phi^{app}$ against $q/2kT$ for the Au/geological MoS$_2$ crystal and (b) the Au/synthetic MoS$_2$ crystal. The solid lines are linear fits to Eq. 8 to obtain the mean Schottky barrier height $\Phi$ and the standard deviation $\sigma$ of the gaussian distribution. (c) Modified Richardson plot for the Au/geological MoS$_2$ crystal and (d) Au/synthetic MoS$_2$ crystal. The two plots are corrected with the $\sigma$ of the gaussian distributions in their respective temperature regimes. Linear fits to Eq. 9 give the $\Phi_4$ from the gradient of the fit and $A_4^{**}$ from the intercept.

**BHEM.** To further verify the presence of a Gaussian distributed SBH, we used ballistic hole emission microscopy (BHEM), which is a nanoscale technique based on scanning tunneling microscopy (STM) to measure the local SBH of the interface for the two different $MoS_2$ crystals. In a typical BHEM experiment, holes are injected into the Au layer from the STM tip by applying a positive tip bias ($V_T$), while the Au layer and the $MoS_2$ substrate are grounded (Figure 1a). Some of the holes travel through the thin Au layer (15 nm) unscattered (the ballistic holes) to reach the metal/semiconductor interface and are collected as the BHEM current ($I_B$) at the Ti/Au ohmic contact if they have enough energy to overcome the Schottky barrier and fulfil the momentum conservation rules [9,34]. To obtain the local Schottky barrier height ($\phi_{BHEM}$), we used the spectroscopy mode by holding the STM tip at a fixed (*x*, *y*) position with the tunnelling current ($I_T$) feedback loop kept on, and collected the BHEM current ($I_B$) as a function of the bias ($V_T$) applied between the tip and the Au layer. We normalized the transmission (*R*) of the interface by taking the ratio of the $I_B$ and the $I_T$ and plotted this ratio against the tip bias to obtain the BHEM spectra (Eq. 10), which is a function of the transmission of the interface to the energy of the electrons. In addition to the main advantage of the nanoscale spatial resolution in BHEM, the *zero bias SBH* (i.e., the equilibrium band alignment) can be measured and visualized directly without having to assume a transport model. For one dataset, we collected approximately 800 BHEM spectra over a 200 nm by 200 nm area and extracted the local SBH by fitting individual spectrum to the Prietsch-Ludeke (P-L) model (Eq. 10) (more details in the Supplemental Information, Figure S2) [34,35].

$$\frac{I_B}{I_T} = R \frac{(\phi_{BHEM} - eV)^{5/2}}{eV} \quad (10)$$

We collected a few datasets over a few different locations on the same sample and Figure 5 shows a representative statistical spread of the local SBH for the two crystals taken from one of the datasets. For the Au/geological MoS$_2$ device, we obtain the local $\phi_{BHEM}$ = 0.86 ± 0.02 eV, while for the Au/synthetic MoS$_2$, we obtained a $\phi_{BHEM}$ = 0.89 ± 0.02 eV. The nanoscale SBH for both the geological and synthetic Au/MoS$_2$ samples are similar, affirming that although the crystal quality is likely to be different, the pristine SBH is similar in value and a gaussian distribution with σ = 0.02 eV is present for both samples.

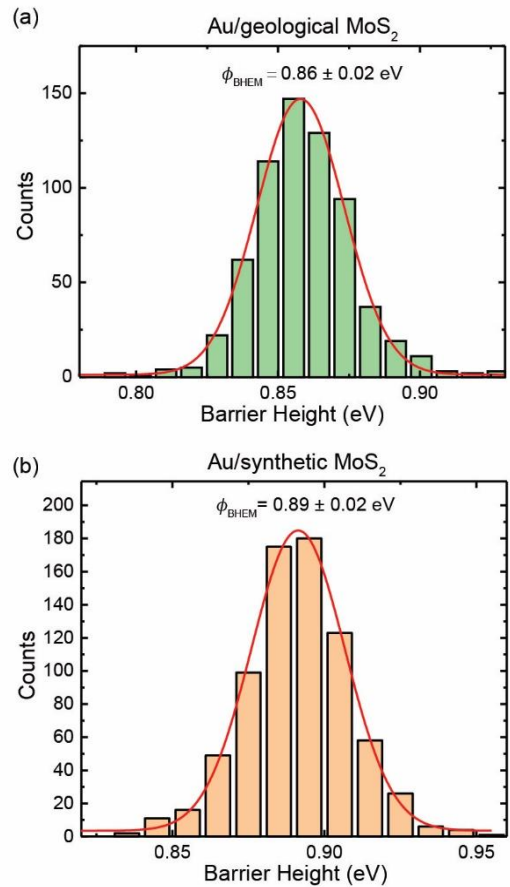

Figure 5. Statistical plot of the Schottky barrier height measured at 120 K for (a) Au (15 nm) /geological MoS$_2$, $\phi_{BHEM}$ = 0.86 ± 0.02 eV, 664 data points. (b) Au (15 nm)/synthetic MoS$_2$, $\phi_{BHEM}$ = 0.89 ± 0.02, 751 data points.

Table 2 shows the summary of the Schottky barrier heights obtained for the Au/geological *p*-MoS$_2$ and Au/synthetic *p*-MoS$_2$ using *I-V-T* measurements and BHEM. We propose that the pristine zero bias SBH for the Au/*p*-MoS$_2$ interface is

~0.86 eV as this value is obtained across complementary techniques and different $MoS_2$ sources. There is at least one defective region present in each sample. We generalize these defects into two regimes, Defect 1 is a lower barrier region, which is present in both the geological and synthetic $MoS_2$ crystals. The density of Defect 1 in the geological crystal could be higher, and thus exists as a shunt region. We propose that the convolution of the lower barrier defect with the pristine SBH can explain the spread of SBH values in the literature as the extracted numbers will differ across different samples and measurement temperatures. The Defect 2 is associated with the higher barrier region possibly by doping, only detected in the geological $MoS_2$ crystal, typically not a dominating effect in the *I-V* measurements due to the exponential relationship of *I* to $\phi$ in the thermionic emission model.

From Table 2, we can conclude that the nanoscale SBH measured using BHEM and macroscale SBH determined using *I-V-T* measurements are in excellent agreement. Standard deviations of the macroscale SBH are one order of magnitude larger than the nanoscale SBH. This can be explained by the non-uniform area of the measurement. *I-V-T* measurements give a weighted average of the SBH across the whole device area (0.196 $mm^2$) which includes contributions from defects such as step edges, impurities, etc. while the nanoscale SBH has a resolution of a few nanometers which measures largely the pristine SBH at the interface. We did not detect the presence of atomic scale defects in BHEM imaging and spectroscopy, i.e. the scanned areas show a single gaussian distributed SBH over hundreds of nanometers. Point defects such as vacancies are not resolved in BHEM measurements likely due to the pinch off effect [36], but they are expected to be present and may be convoluted in the gaussian distributed potential fluctuation of the SBH. This means that a low area density of defects dominates the device in macroscale *I-V-T* measurements and mask the pristine interface typically measured in nanoscale BEEM.

Table 2. Schottky barrier heights of Au/geological *p*-MoS$_2$ and Au/synthetic *p*-MoS$_2$ measured using temperature dependent current – voltage (*I-V-T*) measurements and ballistic hole emission microscopy (BHEM). Defects 1 and 2 denote the two Schottky barrier heights obtained from the double gaussian distribution model of Schottky barrier heights that could arise from different kind of defects.

| | Au/geological MoS$_2$ | | Au/synthetic MoS$_2$ | |
|---|---|---|---|---|
| **Origin of SBH** | Φ (eV) | $A^{**}$ (Am$^{-2}$ K$^{-1}$) | Φ (eV) | $A^{**}$ (Am$^{-2}$ K$^{-1}$) |
| **Defect 1** | Not detected* | - | 0.64 ± 0.08 | 440 000 |
| **Pristine** | 0.88 ± 0.14 | 509 000 | 0.84 ± 0.12 | 407 000 |
| **Defect 2** | 1.17 ± 0.10 | 303 000 | - | - |
| **BEEM** | 0.86 ± 0.01 | - | 0.89 ± 0.02 | - |

*Shunt resistance $R_p$ = 900 kΩ dominates

## VI. Discussion

The different *I-V* behaviors of geological and synthetic MoS$_2$ point to the importance of obtaining high quality crystals for device fabrication. Devices fabricated from synthetic MoS$_2$ show more uniform Schottky diode performances than geological MoS$_2$ as the geological crystal contains many impurities that unintentionally dope the crystal. Although some of the best reported devices are made using geological crystals, the high density of contamination in geological crystals results in poor reproducibility and reliability in the electronic properties of 2D devices. It is difficult to differentiate intrinsic material properties from unwanted dopant effects, which can explain the spread of behaviors of devices in the literature. Addou et al. recently studied the surface of geological MoS$_2$ using STM and showed that the surface of the crystal shows huge variation across the same sample due to impurities [26,27], which suggests that these impurities are important contributors to the electrical behavior of the devices [28]. We have also observed similar defects and impurities in our samples using STM imaging (Supplemental Information, Figure S3), but similar to their analysis, we are unable to to identify the chemical composition of the defects in STM/BEEM and *I-V* measurements as these techniques does not have chemical sensitivity. We have shown that the double gaussian model proposed in this paper can be used to deconvolute the defective SBH from the pristine SBH. This method is also useful for research on engineering ohmic contacts to these materials.

Our MoS$_2$ crystals are *p*-type crystals. In recent literature, most MoS$_2$ crystals reported are *n*-type semiconductors [37–42]. However, early reports have noted that some geological samples are intrinsically *p*-type [43–45] and for synthetic crystals, depending on the growth method, the TMDC crystals can be unintentionally doped n-type or *p*-type due to doping from the chemical transport agent or impurity inclusion [46,47]. The Gaussian distributed model should also work for *n*-type MoS$_2$ and other TMDCs. Cook et al. studied the *n*-type MoS$_2$ using ballistic electron emission microscopy (BEEM) and they obtained the Au/*n*-MoS$_2$ SBH of 0.48 ± 0.02 eV for 16 nm Au layers [48]. Our *p*-type SBH and their *n*-type SBH for ~ 15 nm Au layer sum up to approximately the band gap of bulk MoS$_2$ (~1.34 eV), which provides a useful consistency check, and suggests an unpinned Fermi level. We did not detect any signature of strain in our devices at the limit of resolution of our Raman spectrograph (Supplemental Information, Figure S4), suggesting that the strain did not play a significant role in our measurements. Figure 6 shows the STM image of epitaxial Au films on MoS$_2$ grown using slow deposition method which is an indirect evidence of a clean abrupt interface Au/MoS$_2$ [17,19]. Our results lend support to the presence of an unpinned Fermi level for a well prepared van der Waal's Au/MoS$_2$ interface that is deposited slowly (~0.2 Å/s), consistent with a few recent reports [18,19], and an old photoemission study [49].

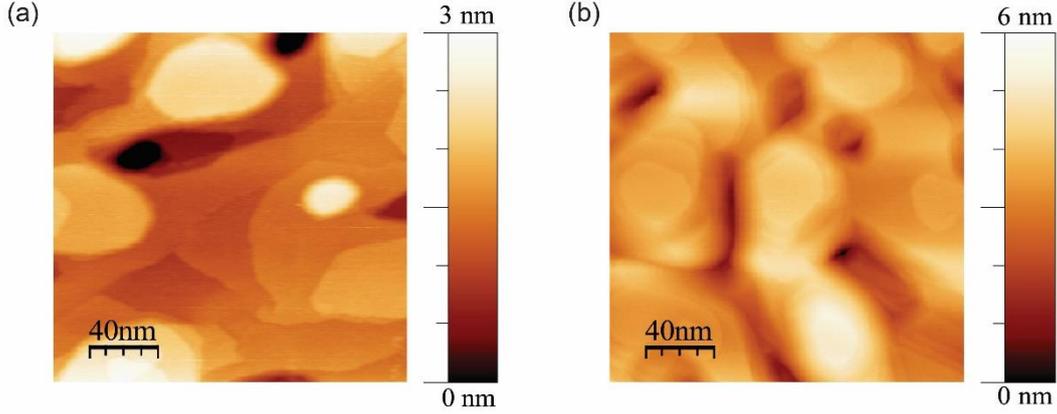

Figure 6: STM images of the Au/MoS$_2$ images at 1.5 V tip bias showing the epitaxial Au overgrowth layer. (a) Au/geological MoS$_2$ (b) Au/synthetic MoS$_2$

We note that the gaussian distributed SBH analysis method we propose in this paper has been applied also to monolayer MoS$_2$ devices [50]. In their report, Moon et al. analyzed the top and edge contact of Au/MoS$_2$ *n*-type FET devices using the gaussian distribution model at different gate bias. They observed the top contact has a larger SBH and larger σ, showing more inhomogeneity than in the edge contact, which has a lower SBH and lower σ. However, they did not use the gaussian corrected Richardson plot, but they indicated that the standard Richardson plot is not valid due to an observed temperature dependence of the SBH. Our method of analysis can be used to bridge the gap between real 2D devices and theoretically proposed models [23] by deconvoluting SBH inhomogeneity from intrinsic material transport behaviors.

### VII. Conclusion

We have shown that the presence of inhomogeneities at the metal/semiconductor interface should be considered in the extraction of device parameters. The analysis of *I-V-T* measurements without considering the gaussian distribution of Schottky barrier heights results in an apparent Schottky barrier height ($\phi^{app}$) which is not reflecting the intrinsic behavior of the interface, but the convoluted effects of low barrier regions (defects), pristine regions and temperature. Using the gaussian modified Richardson plots, reliable mean SBH ($\Phi$) can be extracted, and the extracted effective Richardson constants ($A^{**}$) are close to the theoretical calculated Richardson constants. We report an experimentally measured value of $A^{**}$ = 415 000 ± 85 000 A m$^{-2}$ K$^{-1}$ based on our averaged $A^{**}$ measurements in contrast to $A^{**}$ = 745 000 A m$^{-2}$ K$^{-1}$ typically assumed for p-type MoS2 devices and the Au/p-MoS$_2$ SBH of ~0.86 ± 0.14 eV obtained as an averaged value across two different samples and complementary techniques.

We used BHEM, which is a more tedious but direct method to measure the zero bias SBH without the need to rely on the validity of temperature dependent models, and to experimentally validate the importance of including a gaussian distributed SBH at the nanoscale in conventional *I-V-T* analysis. Our results provide the basic framework for extracting the pristine SBH from temperature dependent *I-V* data and demonstrate that with careful use of the dual parameter ($A^{**}$ + SBH) analysis, we avoid obtaining unphysical numbers that are counter-productive for understanding such interfaces. This implies that the *I-V-T* analysis can yield important insights on the SB inhomogeneities even though it might be a macroscale measurement.

### VIII. Acknowledgements

*We acknowledge funding from the A*STAR Pharos Grant No. 1527000016.*

### IX. Supplemental Information

Description of the Werner method to correct for the high series resistance found in the *I-V-T* curves, comparison of the Prietsch-Ludeke model and the Bell-Kaiser model, impurity scanning tunneling microscope images and strain analysis of our Au/MoS$_2$ using Raman spectroscopy.

Supplemental Information:

# A Gaussian Thermionic Emission Model for Analysis of Au/MoS$_2$ Schottky Barrier Devices

*Calvin Pei Yu Wong,*[1,2,3,*] *Cedric Troadec,*[2] *Andrew T. S. Wee*[1,2,3], *Kuan Eng Johnson Goh,*[2,3*]

[1]*NUS Graduate School for Integrative Sciences and Engineering, National University of Singapore, Centre for Life Sciences, #05-01, 28 Medical Drive, 117456, Singapore*

[2]*Institute of Materials Research and Engineering, A*STAR (Agency for Science, Technology and Research), 2 Fusionopolis Way, #08-03 Innovis, 138634 Singapore*

[3]*Department of Physics, Faculty of Science, National University of Singapore, 2 Science Drive 3, 117551 Singapore*

*Authors to whom correspondence should be addressed: calvin_wong@imre.a-star.edu.sg; kejgoh@yahoo.com

**Supplemental Information**

**Werner method**

In the main text, we mentioned that the Werner method can be used to correct for the high series resistance of the Au/MoS$_2$ diodes. The Werner method is described in detail here in the context of our experimental data.

Under forward bias and with series resistance contribution, the voltage across the diode, $V_d = V - IR_s$ and $V_d \gg kT$, the thermionic emission current (Eq. S1) is given by the simplified form:

$$I_d = I_s \exp\left(\frac{q(V - IR_s)}{nkT}\right) \quad (S1)$$

where $I_d$ is the thermionic diode current. Differentiating Eq. S1 gives the small signal conductance $G = dI_d/dV$ and one obtains:

$$\frac{G}{I_d} = \frac{q}{nkT}(1 - GR_s) \quad (S2)$$

Werner showed that by plotting $G/I_d$ against $G$, named hereafter as the Werner plot, will give a straight line with y-axis intercept of $q/nkT$ where $n$ can be extracted, and x-axis gives the intercept of $1/R_s$.

Figure S1 shows the representative experimental Werner plot for our Au/MoS$_2$ diode at 300 K, 250 K and 200 K from which their respective $n$ and $R_s$ values can be extracted from the y-intercept and x-intercept respectively. After extracting the $R_s$ and $n$ from the Werner plots, the effect of $R$s on the bias voltage can be corrected by subtracting the voltage drop across the series resistance by Kirchhoff's law using $V_D =$

$V - IR_s$. From the $R_s$ corrected *I-V* plots, we fit the standard thermionic emission model (Eq 1.) in the linear diode regime to obtain the Schottky barrier height and the ideality factor. The ideality factors extracted from the Werner plots and the corrected *I-V* plots are within 10% error, validating the Werner method.

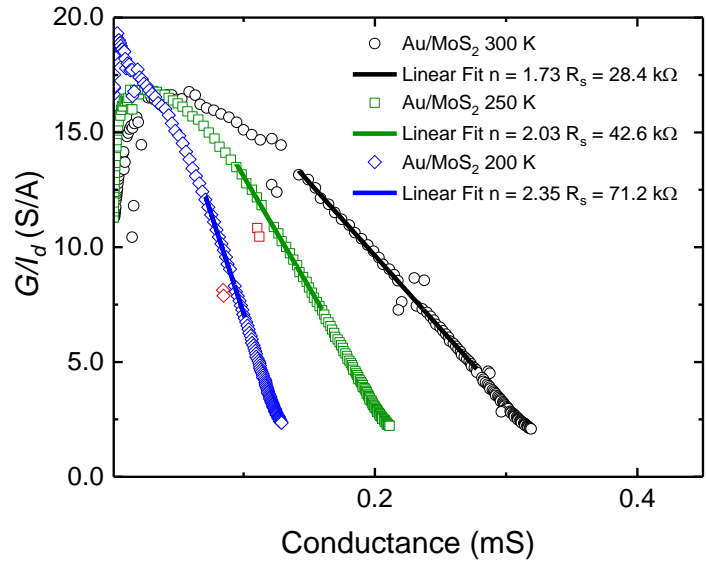

Figure S1. Experimental Werner plot for the Au/geological MoS$_2$ diode at 300 K, 250 K and 200 K. The linear fit to the data according to Eq. 4.7 yields $n = 1.73$ and $R_s = 28.4$ kΩ at 300 K, $n = 2.03$ and $R_s = 42.6$ kΩ at 250 K and $n = 2.35$ and $R_s = 71.2$ kΩ at 200K.

**Prietsch-Ludeke (P-L) model.**

To fit the BHEM spectra, we have considered both the Bell–Kaiser (B-K) [1,2] and Prietsch–Ludeke (P-L) [3] model. The difference in the two models is in the exponent *n* of Equation S5. The B-K model has $n = 2$ while the P-L model is $n = 5/2$. The additional 1/2 power for the P-L model comes from the inclusion of quantum mechanical reflection at the metal/semiconductor interface.

$$\frac{I_B}{I_T} = R \frac{(\phi_B - eV)^n}{eV} \quad (S5)$$

Figure S2 shows that for our experimental data, the P-L model fits better than the B-K model in the fitting range of 0.3 V to 1.3 V, about 0.4 V above the SBH. Therefore, the P-L model is used in our analysis.

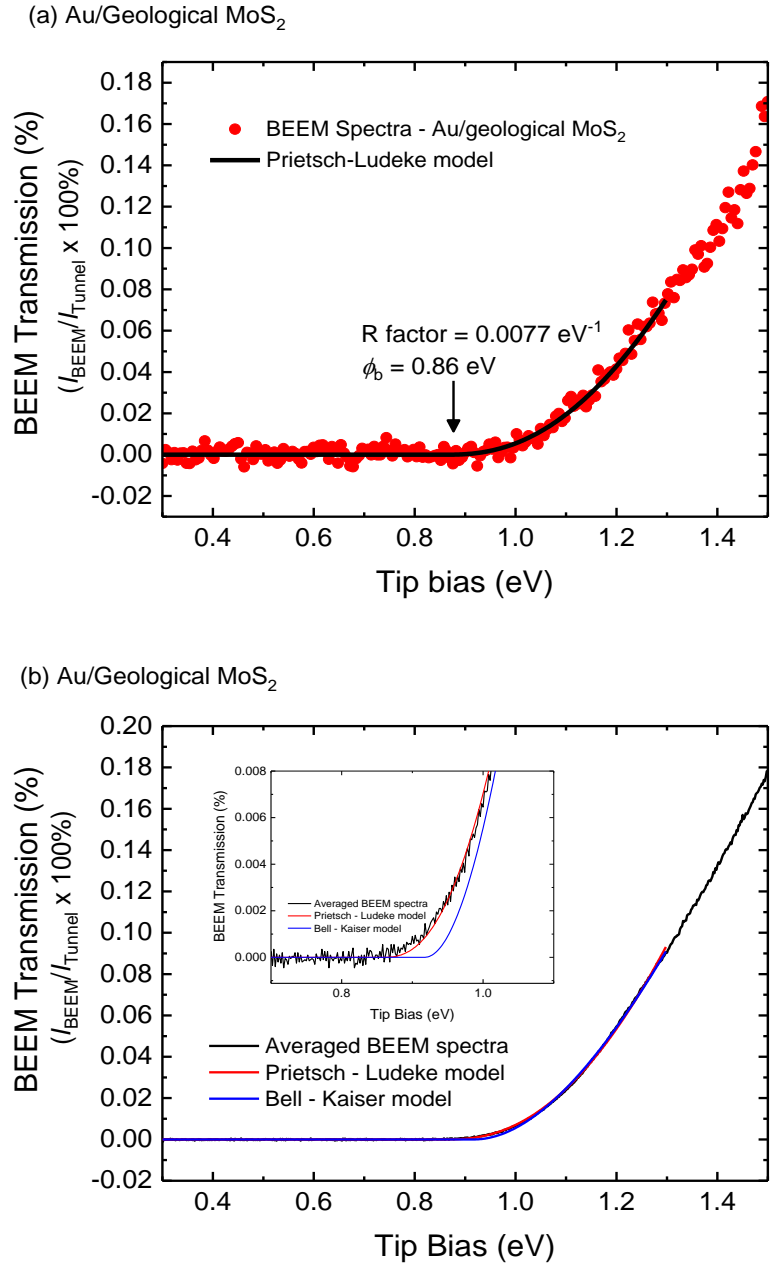

Figure S2 (a) Single representative BEEM spectrum of Au/geological MoS$_2$, fitting the P-L model yields $\phi_B$ = 0.86 eV R-factor = 0.0077 eV$^{-1}$ (b) Averaged BEEM spectrum of Au/geological MoS$_2$ (674 points) to increase signal to noise ratio, spectrum fits better to the P-L model (inset).

**Defect imaging using scanning tunnelling microscopy**

In the main text, we mentioned that we have also seen defects that were previously reported by Addou et al. in our geological MoS$_2$ samples. [4,5] Figure S3 shows the STM images that we obtained from our MoS$_2$ samples. Figure S3a and Figure S3b shows the same region scanned at positive and negative biases respectively. We observed a contrast inversion of some defects where two of the dark defects under positive bias (Figure S3a) appears bright under negative bias conditions (Figure S3b) similar to those defects seen by Addou et al., where they attributed these to metallic impurity clusters present on the surface of the MoS$_2$ crystals. A local depression is also detected in our sample, similar to Addou at al., which could be sulfur vacancies or subsurface sulfur vacancies. These defects could cause the electronic inhomogeneities discussed in the main text of the paper.

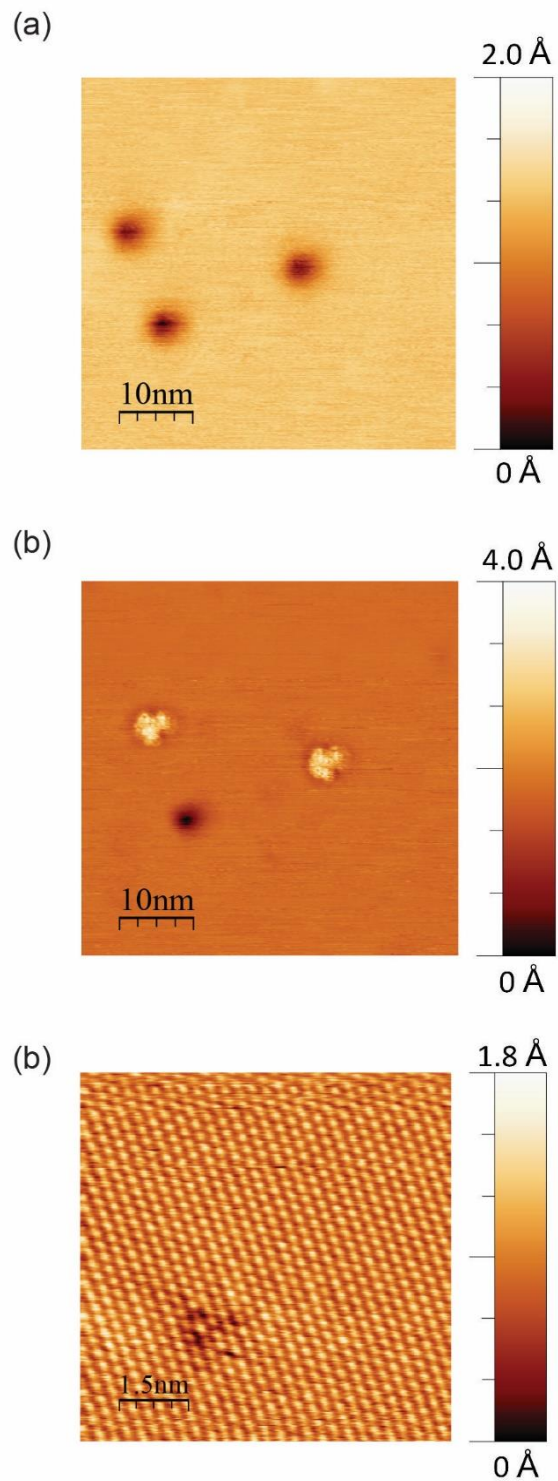

Figure S3. Scanning tunneling microscopy images of (a) the MoS$_2$ surface, metallic cluster like impurities observed at +0.6 V, 2 nA. (b) same area, but at -0.6 V, 2 nA, notice the contrast inversions (c) local depressions observed at -0.6 V, 50 pA

**Strain analysis using Raman spectroscopy.**

To study the effect of strain at our Au/MoS$_2$ interface, we used Raman spectroscopy to compare the relative shifts of the vibrational modes. Figure S4 shows the Raman spectrum obtained on our Au/synthetic MoS$_2$ device using a 532 mn laser at 120 K using a liquid nitrogen flow stage (Linkam). No shift of the Raman peaks was detected when we collected the spectrum on the bare MoS$_2$ surface and on the Au/MoS$_2$ although a decrease in peak intensity can be detected due to attenuation by the 15 nm Au layer. The Raman peaks the all match within 0.2 cm$^{-1}$ which is the resolution limit of our Raman spectrometer. Hence, we can conclude that we did not detect any strain in our Au/MoS$_2$ devices.

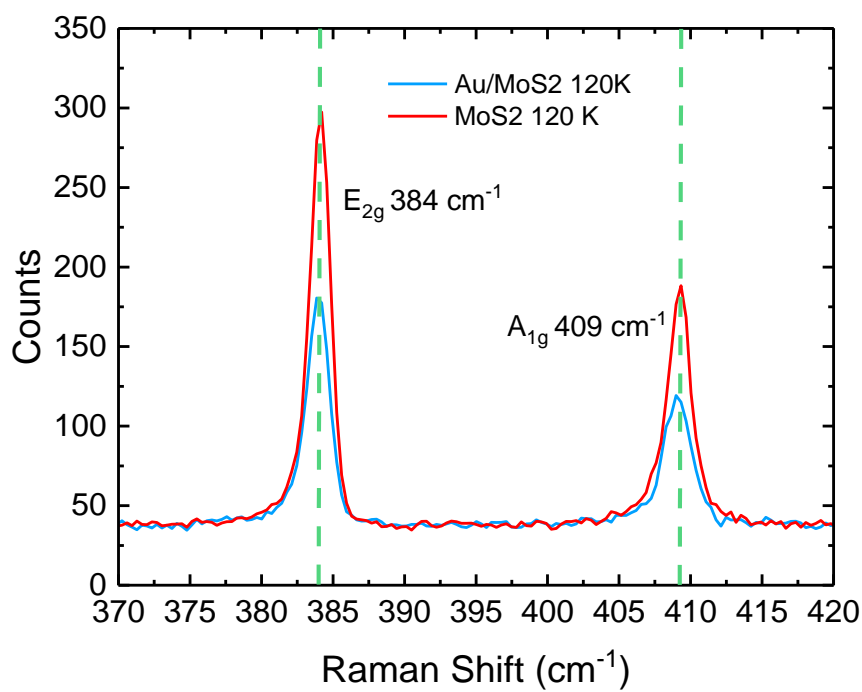

Figure S4. Raman spectrum collected on the Au/MoS$_2$ surface and the bare MoS$_2$ surface at 120 K using liquid nitrogen cooling and 532 nm laser excitation.